\documentclass[12pt]{iopart}

\usepackage{graphicx}
\usepackage{iopams}

\begin{document}

\title {Collective excitations of a dilute Bose gas at finite temperature: TDHFB Theory}

\author{Abdel\^{a}ali Boudjem\^{a}a and Nadia Guebli}

\address{Department of Physics,  Faculty of Exact Sciences and Informatics, Hassiba Benbouali University of Chlef P.O. Box 151, 02000, Ouled Fares, Chlef, Algeria.}
\ead{a.boudjemaa@univhb-chlef.dz}
\vspace{10pt}

\begin{abstract}
Using the time-dependent Hartree-Fock-Bogoliubov approach, where the condensate is coupled with the thermal cloud and the anomalous density, 
we study the equilibrium and the dynamical properties of three-dimensional quantum-degenerate Bose gas at finite temperature.
Effects of the anomalous correlations on the condensed fraction and the critical temperature are discussed.
In uniform Bose gas, useful expressions for the Bogoliubov excitations spectrum, the first and second sound, the condensate depletion 
and the superfluid fraction are derived.
Our results are tested by comparing the findings computed by Quantum Monte Carlo simulations.
We present also a systematic investigation of the collective modes of a Bose condensate confined in an external trap.
Our predictions are in qualitative agreement with previous experimental and theoretical results.
We show in particular that our theory is capable of explaining the so-called anomalous behavior of the $m=0$ mode.

\end{abstract}

%

\section{Introduction}
The study of collective mode oscillations of a Bose-Einstein condensate (BEC) in a dilute gas has received much attention during the last 20 years.
In fact, elementary excitations and collective modes provide important and precise information  on the quantum many-body nature of these systems and on the role of interactions.
Experimentenlly, low-lying collective excitations have studied over a range of temperatures \cite{Kitt, Jin,Mew, Jin1, Tilm, Haller, Bess, Andr}. 
Zero-temperature findings have agreed well with predictions based on a mean-field description of the weak interatomic interactions \cite {Jin1,Edw, String, Castin}. 

At finite temperature, the excitations of Bose gas have been studied theoretically using the Hartree-Fock-Bogoliubov (HFB) Popov approximation but 
this agrees with experiment only at temperatures $T< 0.6\,T_c$ \cite {Hut}. 
Good agreement for the full temperature rangem for the $m=2$ mode was obtained using the generalized HFB (GHFB)
approach which includes the anomalous average \cite {Hut}. However, both approaches were unable
to explain the upward shift (often known as an "anomalous behavior") of the $m = 0$ mode.
A possible explanation for such an anomalous behavior of the $m=0$ mode was given by Stoof's group \cite{Stoof1, Stoof2} 
in terms of a crossover from out-of-phase to in-phase oscillations of the condensed and noncondensed atoms at high temperature.
Simulations \cite{Zar3} based on the Zaremba-Nikuni-Griffin (ZNG) theory which includes semiclassically the thermal cloud dynamics (Boltzmann equation) \cite{Zar,Zar1},
and on the second-order quantum field theory \cite{Morgan} give good quantitative agreement for both modes. 
However, none of these studies took into account the full dynamics of the anomalous average. 
This latter plays a crucial role, especially near a Feshbach resonance \cite{Morg, Holland, Koh}. Furthermore, the absence of the pair anomalous density
renders the system unstable \cite{Boudjbook}.

In this article, we use an alternative approach to study the dynamics of a trapped BEC at finite temperature
called time dependent Hartree-Fock-Bogoliubov (TDHFB) \cite {Boudjbook, Boudj, Boudj1,Boudj2,Boudj3,Boudj4,Boudj5,Boudj6,Boudj7, Boudj8, Boudj12}.
The TDHFB theory is a set of coupled nonlinear equations of motion for the condensate, thermal cloud and the anomalous density. 
Our formalism which ensures the conservation laws, and a gapless spectrum of collective excitations, is based on the time-dependent 
Balian-V\'er\'eroni (BV) variational principle \cite{Balian}.
This latter optimizes the evolution of the state according to the relevant observable in a given variational space \cite{Boudj1,Balian}.
The BV variational principle offers an elegant starting point to build approximations of the many-body dynamics and has been successfully applied to a wide variety of problems. 
The equation of motion for the condensate provided by our theory is closest in form to that obtained in 
\cite{Hut, Zar, Griff, Prouk, Tim, Gard, Stoof, Wals,  Giorg1, Holl, Cherny, Bul}. 
However, in contrast, our equation for the noncondensed density has no analogue in the literature. 
The TDHFB equations permit us to deeply investigate the behavior of Bose condensed gases at nonzero temperature, which involves interactions 
between the condensate and the thermal cloud. They give us also insightful discussions of the temperature dependence 
of the collective mode frequencies. 
It is therefore, particularly essential to use our TDHFB theory to explain the JILA experiment and to illustrate the anomalous behavior of the $m=0$ mode.


The rest of the paper is organized as follows. In section \ref{flism}, we introduce the TDHFB model. 
Accordingly, coupled equations governing the dynamics of the condensate, noncondensate and the anomalous component are derived.
We present numerical results for the temperature dependence of the condensed fraction and for the shift of the critical temperature 
of a condensate confined in a cylindrically symmetric harmonic trap. 
We compare our predictions with available  theoretical treatments and quantum Monte carlo (QMC) simulations.
In section \ref{CollExc}, starting from the TDHFB-de-Gennes equations (TDHFBdG) equations we examine effects 
of the temperature and  anomalous correlations on the collective modes of a weakly interacting Bose gas.
In the homogeneous case,  we calculate in particular corrections induced by the anomalous terms to the first and second sound, 
the thermal depletion and the superfluid fraction (see Sec.\ref{Homg}).  Our results are verified through comparison with QMC simulations.
Section \ref{CM} deals with the collective frequencies of the condensate taking into account the dynamics of the thermal component
in a trap corresponding to the JILA experiment \cite{Jin1}. 
We compare our findings with previous experimental data and theoretical treatments such as the GHFB \cite{Hut, Hut1} 
and the ZNG theory \cite{Zar3}. 
Conclusions and discussions remain for section \ref{concl}.

\section{Formalism} \label{flism}

We consider an atomic trapped gas of $N$ bosons interacting via the short range potential.
The TDHFB equations which we choose to employ here may be represented  as 
\cite {Boudjbook, Boudj, Boudj1,Boudj2,Boudj3,Boudj4,Boudj5,Boudj6,Boudj7, Boudj8,Boudj12}:
\begin{eqnarray}
&i\hbar \frac{\partial \Phi }{\partial t} = \left [-\frac{\hbar^2}{2m} \Delta +V+g_{ca} n_c+2g\tilde{n} \right]\Phi, \label{cond} \\ 
&i\hbar \frac{\partial \tilde{m} }{\partial t} = 4\left[-\frac{\hbar^2}{2m} \Delta+V+G (2\tilde {n}+1)+2gn\right]\tilde{m}, \label{anom}
\end{eqnarray}
where  $V({\bf r})$ is the trapping potential, 
$\Phi ({\bf r})=\langle \hat\psi ({\bf r})\rangle$  is the condensate wavefunction, $n_c({\bf r})=|\Phi({\bf r})|^2$ is the condensed density,
the noncondensed density $\tilde{n}({\bf r})$ and the anomalous density $\tilde{m}({\bf r})$ are identified respectively with 
${\langle \hat\psi^{+} ({\bf r}) \hat\psi ({\bf r})\rangle }-\Phi ^{*} ({\bf r}) \Phi ({\bf r})$ 
and ${\langle \hat\psi{ ({\bf r})} \hat\psi ({\bf r})\rangle}-\Phi {({\bf r})}\Phi {({\bf r})}$, where $\hat\psi^{+}$ and $\hat\psi$ 
are the boson destruction and creation field operators, respectively. The total density in BEC is defined by $n=n_c+\tilde {n}$.
The renormalized coupling constant $g_{ca}$ is defined as  \cite{Hut, Boudj2, Wrig}
\begin{equation}  \label{RCC}
g_{ca}=g \left(1+ \frac{\tilde {m}}{\Phi ^2} \right).
\end{equation}
This spatially dependent effective interaction allows us to avoid the well known issues supervening from the inclusion of the anomalous density 
such as ultraviolet divergences and the appearance of an unphysical gap in the excitations spectrum. 
The parameter (\ref{RCC}) is equivalent to the many body $T$-matrix derived in \cite{Hut, Boudj2}. 
The second term in $g_{ca}$ describes the change in the correlation, i.e. pairing, between atoms
which produced by Bogoliubov terms in the Hamiltonian. Pairs of this sort become bound in the case of Fermi gases below the BCS transition temperature. 
The parameter $G$ is related to $g_{ca}$ via $G=gg_{ca}/4(g_{ca}-g )$ with $g=4\pi \hbar^2a/m$ and $a$ is the $s$-wave scattering length.


In our formalism the noncondensed and the anomalous densities are not independent. 
By deriving an explicit relationship between them, it is possible to eliminate $\tilde{n}$ via \cite{ Boudj3, Boudj5, Cic,Cherny}:
\begin{equation}  \label{Inv}
\tilde{n} = \sqrt{|\tilde{m} |^2+\frac{1}{4}I}-\frac{1}{2}I,
\end{equation}
where $I$ is often known as Heisenberg invariant \cite{Cic} and represents the variance of the number of noncondensed particles.
One can easily check that Eq.(\ref {Inv}) can reproduce the mean-field result based on the HFB approximation.
Working in the Bogoliubov quasiparticles space one has 
$\hat a_{\bf k}= u_k \hat b_{\bf k}-v_k \hat b^\dagger_{-\bf k}$, where $\hat b^\dagger_{\bf k}$ and $\hat b_{\bf k}$ are operators of elementary excitations
and $ u_k,v_k$  are the standard Bogoliubov functions. In the quasiparticle vacuum state, $\tilde{n}$ and $\tilde{m}$ may be written as
$\tilde{n}=\sum_k \left[v_k^2+(u_k^2+v_k^2)N_k\right]$ and $\tilde{m}=-\sum_k \left[u_k v_k (2N_k+1)\right]$,
where $N_k=[\exp(\varepsilon_k/T)-1]^{-1}$ are occupation numbers for the excitations and 
$ u_k,v_k=(\sqrt{\varepsilon_k/E_k}\pm\sqrt{E_k/\varepsilon_k})/2$ are the Bogoliubov functions with $E_k=\hbar^2k^2/2m$ being the energy of free particle. 
Utilizing the orthogonality and symmetry conditions between the functions $u_k$, $v_k$ and using the fact that $2N (x)+1= \coth (x/2)$, we obtain \cite{Boudj7,Boudj8}:
\begin{equation}\label {heis}
I =(2\tilde{n} +1)^2-4|\tilde{m} |^2 = \sum_k \coth^2\left(\frac{\varepsilon_k}{2T}\right) .
\end{equation}
Expression (\ref{Inv}) clearly shows that $\tilde{m}$ is larger than $\tilde{n}$ at zero temperature, so the neglect of the anomalous density 
is an unjustified approximation.

At very low temperature where $I \rightarrow 1$, equation (\ref{Inv}) may be expanded in a Taylor series as 
\begin{equation}  \label{Tay}
\tilde{n}=|\tilde{m}|^2-|\tilde{m}|^4+2|\tilde{m}|^6-\cdots
\end{equation}
Let us now multiply (\ref{anom}) by $\tilde{m}^*$ and subtract  the complex conjugate of the resulting expression. 
After having taking only the leader term in the expansion (\ref{Tay}), we obtain
\begin{equation} \label{nonc}
i\hbar \frac{\partial  \tilde{n}}{\partial t} = 4\left[-\frac{\hbar^2}{2m} \Delta+2V+g_{ac} \tilde {n}+4gn_c +2G\right] \tilde{n} , 
\end{equation}
where $g_{ac}=4(G+g)$.\\
Equations (\ref{cond}), (\ref{Tay}) and  (\ref{nonc})  are the closed set of the TDHFB equations for a condensate
coexisting in a trap with a thermal cloud and a pair anomalous density. 
The main advantage of our equations is that they do not neglect the effect of the anomalous correlation.
Nonetheless, the role of the anomalous terms is implicitly taken into account in the renormalized coupling constants $g_{ac}$.
Furthermore, the theory is valid for arbitrary interacting Bose systems, whether equilibrium or nonequilibrium, homogeneous or inhomogeneous.
Another important feature of the TDHFB equations is that they treat the dynamics of the condensate 
and the thermal cloud on an equal footing and in a self-consistent way which is not the case 
in the aforementioned methods such as the standard HFB approximation and the ZNG approach.
In the former,  the condensate is assumed to move in a static thermal cloud while in the latter (i.e. the ZNG), 
the condensate and the noncondensate are looked in qualitatively different ways
as we have stated above. 
In addition, although the TDHFB equations have been derived for a trapped gas at finite temperatures, 
they remain valid at  $T =0$, since $\tilde {n}$ becomes negligible, the condensate equation (\ref{cond})
reduces to the standard Gross-Pitaevskii equation of a pure condensate at $T=0$. 
One should stress that other useful versions of the TDHFB theory have been derived in \cite{Prouk,Tim, Gard, Stoof, Wals, Giorg1, Holland, Cherny, Bul,Yuk1} 
using different techniques (for more details, we refer the reader to \cite {Boudjbook}).

The total number of particles is defined as $N=N_c+\tilde N$, where
\begin{equation} \label{Nprt}
N_c=\int  d{\bf r} \,n_c ({\bf r}) ,   \,\,\,\,\,\,\,\,\,\,\,\,\,\,\,\,\,   \tilde N=\int  d{\bf r}\, \tilde{n} ({\bf r}),
\end{equation}
 are, respectively, the number of condensed and noncondensed particles.

Before studying the collective modes of a Bose condensate, it is necessary to elucidate the behavior of the condensed fraction and  the critical temperature.
We focus in particular on the peculiar effects of the anomalous fluctuations.
The trap potential is modelled as
\begin{equation} \label{extpot}
V({\bf r})=\frac{1}{2}m \omega_{\rho}^2 (\rho^2+\lambda^2 z^2),
\end{equation} 
where $\rho^2=x^2+y^2$, and $\lambda=\omega_{z}/\omega_{\rho}$ is the ratio between the trapping frequencies in the axial and radial directions.

We solve the static equations (\ref{cond})-(\ref{nonc}), which constitute a new fashion of the HFB equations, numerically (see details in Appendix). 
The numerical calculation of our equations is relatively easy and is not time-consuming 
even for large numbers of particles compared to the standard HFB equations \cite{Boudj7}. 
These latter become rapidly unstable for higher modes and for increasing temperatures \cite{Hut}. 

\begin{figure} [ htb] 
  \centering
  \includegraphics[scale=1] {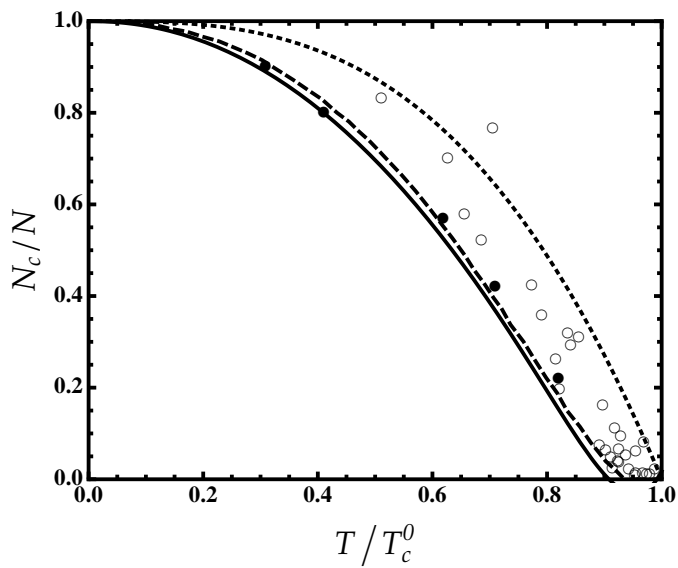}
  \caption{Condensed fraction as a function of reduced temperature $T/T_c^0$ (where $T_c^0$ is the ideal gas critical temperature). 
   Solid line: our predictions for $\eta=0.4$. Dashed line: the results of the HFB-Popov approximation  for $\eta=0.4$ \cite {Giorg, Franc}.
  Open circles: the experimental data of \cite{Jin}, corresponding to $\eta$ in the range 0.39-0.45.
  Solid circles: path-integral QMC results of \cite{Krauth}  with $\eta=0.35$. Dotted line: the noninteracting  case $\eta=0$ and $N_c/N=1-(T/T_c^0)^3$.}
   \label{CF}
\end{figure}

Our simulation parameters are chosen to match those of the experiment \cite{Jin}, where  $\omega_{\rho} = 2\pi \times 129$ Hz, 
$\lambda= \sqrt{8}$ and the $s$-wave scattering length $a=110 a_0$ ($a_0$ is the Bohr radius).
In what follows, we express lengths and energies in terms of the transverse harmonic oscillator length $l_0=\sqrt{\hbar/m \omega_{\rho}}$
and the trap energy $\hbar \omega_{\rho}$, respectively. 
We introduce a useful dimensionless parameter $\eta=1.57 (N^{1/6} a/l_0)^{2/5}$ \cite{Franc}.
The experimental data of \cite{Jin} corresponding to $\eta$ in the range 0.39-0.45. 

In figure \ref{CF} we compare our predictions for the condensed fraction $N_c/N$ with the HFB-Popov calculation of \cite {Giorg,Franc},
path integral QMC simulation of \cite{Krauth}, experimental data of  Ref \cite{Jin}  and the noninteracting gas.
As is clearly seen, our results, the HFB-Popov \cite {Giorg,Franc} and QMC  \cite{Krauth} curves  agree at very low temperature.
At $T\geq 0.2 T_c$, there is a distinct difference between our findings and the other approaches
which we attribute to the influence of the anomalous fluctuations \cite {Hut,Boudj3, Boudj4}. 
We see also that our theory gives an even better agreement crossing experimental data at temperature close to the transition $T\sim T_c$.

Evidently, when the number of atoms in the trap increases as e.g. in Na atoms experiment \cite{Mew}, the effects due to the interaction 
become more and more significant, and thus the impacts of the anomalous correlations become substantial.
We may infer that such anomalous effects which arise from the interactions \cite{Griff} may reduce both the critical temperature and the condensed fraction
in agreement with exact QMC methods \cite{Krauth, Holtz}.
In our formalism the shift of the critical temperature $T_c$ can be determined via (\ref{Inv}) where 
$I_{\bf k} ({\bf r}) =\coth^2\left( {\cal E}_k ({\bf r})/2T_c\right)$ with ${\cal E} ({\bf r})$ being the energy of the system.

\section{Collective excitations} \label {CollExc}

Our aim in this section is to examine effects of the temperature and anomalous correlations on the collective modes of a weakly interacting Bose gas.

\subsection{TDHFB-de-Gennes equations} \label {BdGE}

We turn our attention now  to deriving the BdG equations by linearizing the Eqs.(\ref {cond}) and  (\ref{nonc})  using the decomposition
$\Phi=(\Phi_0+\delta \Phi) e^{-i\mu_ct/\hbar}$ and $\tilde n=(\tilde n_0+\delta \tilde n) e^{-i\mu_nt/\hbar}$ \cite{Giorg1}, 
where $\mu_c$ is the chemical potential of the condensate,
$\mu_n$ is the chemical potential of the noncondensate, $\delta \Phi$ and $\delta \tilde n$ 
are respectively, small fluctuations of the condensate and the thermal cloud around their equilibrium values.  We then look for the solution of the form
\begin{eqnarray}  
&\delta \Phi= u_j^c( {\bf r}) e^{-i\epsilon_j t/\hbar}+v_j^c( {\bf r}) e^{i\epsilon_j t/\hbar},  \label{FLC} \\
&\delta \tilde n= u_j^n( {\bf r}) e^{-i\epsilon_j t/\hbar}+v_j^n( {\bf r}) e^{i\epsilon_j t/\hbar},  \label{FLN}
\end{eqnarray}
where $u\,,v$ are normalized according to $\int d {\bf r} [u^2( {\bf r})- v^2({\bf r})]=1$.
Rather than dealing with the functions $u_j$ and $v_j$, it is useful to define the functions $f_j^{\pm}=u_j\pm v_j$ which are 
solutions of the TDHFBdG equations:
\begin{eqnarray} 
\varepsilon_j f_j^{c-}({\bf r})&=\left[{\cal L}_c+g_{ca} n_c+2g\tilde {n}\right] f_j^{c+}({\bf r}), \label{B1:td} \\
\varepsilon_j f_j^{c+}({\bf r})& =\left[{\cal L}_c+3g_{ca} n_c+2g\tilde {n}\right] f_j^{c-}({\bf r}) +4g\sqrt{\tilde {n} \, n_c}  f_j^{n-}({\bf r}), \label{B2:td}   \\  
\varepsilon_j f_j^{n-}({\bf r})&=4\left[{\cal L}_n+g_{ac} \tilde {n}+4g n_c \right] f_j^{n+}({\bf r}), \label{B3:td} \\ 
\varepsilon_j f_j^{n+}({\bf r})&= 4\left[{\cal L}_n+2g_{ac} \tilde {n}+4g n_c \right] f_j^{n-}({\bf r}) +16g \sqrt{\tilde {n}\,n_c}  f_j^{c-}({\bf r}),  \label{B4:td}
\end{eqnarray}
where ${\cal L}_c=-(\hbar^2/2m) \Delta + V -\mu_c$ and ${\cal L}_n=-(\hbar^2/2m) \Delta + 2V -\mu_n+2G$.\\
Equations (\ref{B1:td})-(\ref{B4:td}) form a complete set to calculate the ground state and the collective modes of the condensate and the thermal cloud 
of a Bose gas at  low temperatures.

\subsection{Homogeneous case} \label {Homg}
In the homogeneous case where $V(r)=0$, the chemical potentials read
\begin{eqnarray}
&\mu_c=g_{ca} n_c+2g\tilde {n},  \label{chimB1} \\
&\mu_n=4\left (g_{ac} \tilde {n}+4g n_c +2G\right).  \label{chimm1}
\end{eqnarray}
A lengthy but straightforward diagonalization of the BdG equations (\ref{B1:td})-(\ref{B4:td}) yields the following two-branch dispersion relation of
the coupled modes of the condensate and the thermal cloud
\begin{equation} \label{dis}
\varepsilon_k^{\pm2}= \frac{{\varepsilon_k^c}^2+{\varepsilon_k^n}^2 } {2} \pm \sqrt{\frac{ \left( {\varepsilon_k^c}^2-{\varepsilon_k^n}^2\right)^2 } {4}+ 64 E_k^2 g^2n_c\tilde {n} },
\end{equation} 
where $\varepsilon_k^c=\sqrt{E_k^2+2g_{ca} n_c E_k }$, is the standard single condensate Bogoliubov dispersion relation,
$\varepsilon_k^n=4\sqrt{E_k^2+g_{ac} \tilde{n} E_k }$ is the thermal cloud dispersion relation,  and $E_k=\hbar^2k^2/2m$ is the energy of free particle.\\
Expression (\ref{dis}) means that the dynamics and the thermodynamics of the condensate and the thermal cloud mixture 
are essentially the dynamics and thermodynamics of two sets of noninteracting quasi-particle with energies $\varepsilon_k^{\pm}$.

In the limit $k \rightarrow 0$,  we have $\varepsilon_k^c= \hbar c^c k$ and  $\varepsilon_k^n= 2\sqrt{2}\hbar c^n k$ 
where $c^c=\sqrt{n_cg_{ca}/m}$ and $c^n=\sqrt{\tilde {n} g_{ac}/m}$ are the sound velocities corresponding to the condensate and the noncondensate, respectively. 
Hence, the total dispersion gives two phonon solutions
\begin{equation} \label{sound}
\varepsilon_k^{\pm}= \hbar c^{\pm} k,
\end{equation} 
where the sound velocities are given by
\begin{equation} \label{sound1}
c^{\pm2} =\frac{1}{2} \left[ {c_c}^2+{c_n}^2 \pm \sqrt{ \left( {c_c}^2-{c_n}^2\right) ^2 + 256 \frac{g^2}{ g_{ca}\, g_{ac} } c_c^2 c_n^2} \right].
\end{equation} 
The $\varepsilon_k^+$ mode in (\ref{sound}) clearly corresponds to first sound while  $\varepsilon_k^-$  is the second sound mode.
The sound velocities in (\ref{sound1})  can be expressed in terms of condensed, noncondensed and anomalous fractions  through the above definitions of $g_{ca}$ and $g_{ac}$ as 
\begin{equation} \label{sound2}
c^{\pm2} =\frac{c_0^2}{2} \left[ 1+\gamma_1 \pm \sqrt{ (1+\gamma_2)^2 + 1024 \gamma_3} \right],
\end{equation} 
where $c_0=\sqrt{n_cg/m}$ is the standard sound velocity and the parameters $\gamma_i$
$$\gamma_1=\frac{\tilde{m}}{n_c}+ \frac{8\tilde{n}} {n_c} +\frac{4\tilde{n}} {\tilde{m}},$$
$$\gamma_2=\frac{\tilde{m}}{n_c}- \left(\frac{8\tilde{n}} {n_c} +\frac{4\tilde{n}} {\tilde{m}} \right),$$
$$\gamma_3=\frac{3\tilde{n}}{n_c}+ \frac{2\tilde{m} \tilde{n}} {n_c^2} +\frac{\tilde{n}} {\tilde{m}},$$
represent the temperature and the interaction corrections to the sound velocities.
It is important to stress that the noncondensed and the anomalous fractions can be evaluated through (\ref{Inv}).
Equations (\ref{sound2}) which take into account higher order terms in $g$ and effects of the anomalous density, 
constitute an exciting new extension of the first and second sound velocities obtained by the ZNG theory \cite{Zar2}.
Importantly, Eqs.(\ref{sound2}) show that the velocity of the two modes has the tendency to cross at temperatures close zero.
This phenomenon which called hybridization, was first investigated by Lee and Yang  \cite{Lee}.
For $g^2> g_{ca} \,g_{ac}$ or equivalently $\tilde{m}/ n_c  \in  [-1, (3-\sqrt{17})/4]$, the second sound $c^{-} $ becomes negative indicating that the
system is unstable against long wavelength excitations. 
This means that $c^{-} $ strongly involves oscillation of the condensate atoms (superfluid density) and displays a soft mode in the normal fluid \cite{Zar2}.
The relationship between our sound velocities and the standard Landau's equations can be found using the thermodynamic relations of an ideal Bose gas model.

\begin{figure} [ htb] 
  \centering
  \includegraphics[scale=0.9] {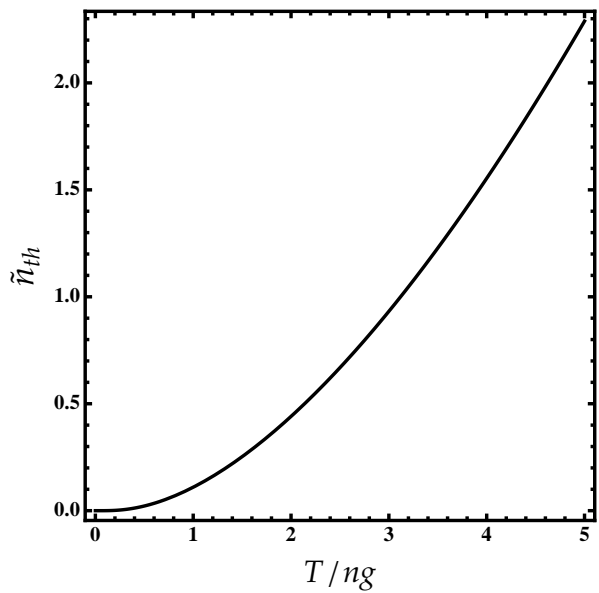}
  \caption{Thermal depletion from Eq.(\ref{dep}) as a function of $T/n_c g$ for $g^2/ g_{ca} \,g_{ac}=0.5$.}
   \label{Thd}
\end{figure}

In what follows we consider only $\varepsilon_k^{+}$ since it is the relevant branch regarding stability.\\
Assuming that $\tilde {n}= |\tilde {m}|^2$, the thermal depletion can be computed via (\ref{heis}) and (\ref {Tay}) as
\begin{equation}  \label{dep} 
\tilde {n}_{th}=\int \frac{d {\bf k}}{(2\pi)^3} \frac{1}{2\sinh (\varepsilon_k/2T)}. 
\end{equation}
Obviously, Eq.(\ref{dep}) is different from the standard expression of $\tilde {n}_{th}$ signaling that
the anomalous correlations may enhance the thermal depletion. \\
At low temperatures, $T\ll gn_c $, the main contribution to integral (\ref{dep}) comes from the region of small momenta 
($\varepsilon_k \equiv \varepsilon_k^+= \hbar c^+ k$).
We then obtain
\begin{equation}  \label{dep1}
\frac{\tilde {n}_{th}}{n_c}= {\cal F}(T) \frac{\tilde {n}_{T}}{n_c},
\end{equation}
where 
$$ {\cal F}(T) =\frac{84\,\zeta (3) }{\pi^2} \left(1+\frac{\tilde {m}}{n_c} \right)^{-3/2}  \left(\frac{c^c}{c^+}\right)^3 \frac{T}{n_c g},$$
$\zeta (3)$ is the Riemann Zeta function,
and
\begin{equation}  \label{depT0}
\frac{\tilde {n}_T}{n_c} =\frac{2}{3} \sqrt{\frac{n_c a^3}{\pi}}\left(\frac{\pi T}{n_c g}\right)^2
\end{equation}
is the thermal depletion obtained using the standard Bogoliubov theory \cite{Boudj9}.\\
We see from Eq.(\ref{dep1}) that the thermal contribution of the noncondensed density is $\propto T^3$
in disagreement with the Bogoliubov depletion (\ref{depT0}) where $\tilde {n}_T \propto T^2$.  
At temperatures $T\ll gn_c $ and for $\tilde {m}/n_c \ll1$, $\tilde {n}_{th}$ is much smaller than $\tilde {n}_T$.
At $T \gg gn_c$, it is clear that the noncondensed density agrees with the noncondensed density of an ideal Bose gas since $\varepsilon_k=E_k$ and $\tilde {m} =0$.
The thermal depletion increases monotonically with temperature as is seen  in figure \ref{Thd}.

The quantum depletion can be obtained through (\ref{heis}):
\begin{equation}  \label{Qdep}
\tilde {n}= \frac{1}{6\sqrt{2}\pi^2}  \left(\frac{mc^+} {\hbar} \right)^3. 
\end{equation}
In the dilute limit where $n_ca^3 \ll 1$ and $\tilde m/n_c \ll 1$, the noncondensed density reduces to that obtained earlier within the Bogoliubov approach
namely: $\tilde {n} \simeq (8/3) n_c \sqrt{n_c a^3/\pi}$. 
Therefore, one can expect that the impact of quantum depletion onto the structure of the low-energy spectrum is not important. 
This is indeed in contrast with the recent result of \cite {Beink}, where it has been found that the depletion of  BECs in one-dimensional triple well optical lattices 
of only around 1\% leads to clear deviations from the BdG spectra.   
We should note at this stage that the dimensionality of optical lattices may significantly influence the behavior of the quantum depletion \cite {Bendach} 
and thus, modifies the Bogoliubov excitations.

\begin{figure}   [ htb] 
  \centering
  \includegraphics[scale=0.9] {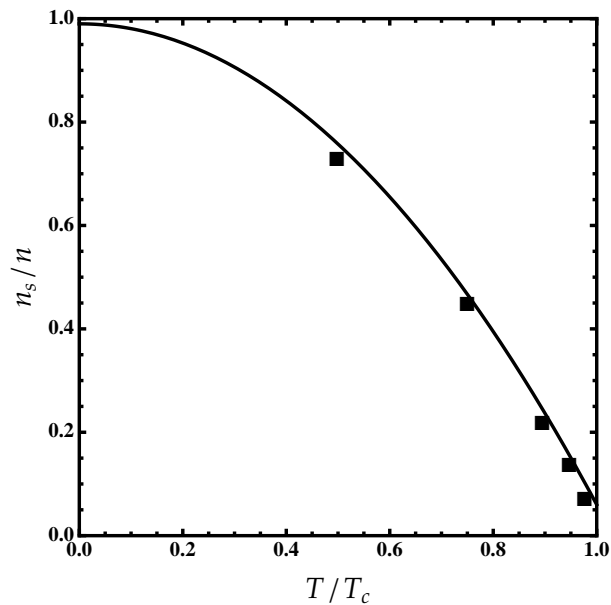}
  \caption{ Superfluid fraction as a function of reduced temperature for $na^3=10^{-6}$.  Solid line: our predictions from Eq.(\ref{supflui1}). 
Squares: QMC results of \cite{Cap}. The transition temperature is given by $T_c=T_c^0 (1+ C_0 an^{1/3})$, where $C_0 = 1.29 \pm 0.05$ \cite{Cap}.}
 \label{SpF}
\end{figure}

The superfluid fraction is defined as (c.f. \cite{ Boudj5, LL9, Boudj10})
\begin{equation}\label {supflui}
\frac{n_s}{n} =1-\int E_k\frac{\partial N_k}{\partial\varepsilon_k}\frac{d^3k}{(2\pi)^3} \frac{}{},
\end{equation}
where $N_k=[\exp(\varepsilon_k/T)-1]^{-1}$ are occupation numbers for the excitations.  \\
Again at $T\ll gn_c$, a straightforward calculation leads to 
\begin{equation}\label {supflui1}
\frac{n_s}{n} =1-\frac{1}{45} \sqrt{\frac{n_c a^3}{\pi}} \left(\frac{2\pi T}{n_c g}\right)^4 \left[ \frac{2} {1+\gamma_1 + \sqrt{ (1+\gamma_2)^2 + 1024 \gamma_3} }\right]^{5/2}.
\end{equation}
The subleading term in Eq.(\ref {supflui1}) represents our correction to the superfluid fraction.
Remarkably, at $T=0$, the whole liquid is superfluid and $n_s=n$.

In figure \ref{SpF},  we compare QMC results of \cite{Cap} for the superfluid fraction at temperatures $T \leq T_c$, with our theoretical prediction 
from Eq.(\ref{supflui1}). 
We observe that the analytical expression deviates from the QMC predictions at temperatures in the range $0<T<0.7\,T_c$. 
This discrepancy arises from the fact that the thermal normal and anomalous fluctuations contribution to the superfluid fraction is important.
At $T>0.7\,T_c$,  the two curves (analytic and numeric) agree with each other since the anomalous fluctuations vanish at $T\sim T_c$ \cite{Boudj2,Griff}.

\subsection {Trapped gas}\label {CM}

Here we analyze  the collective oscillations of interacting Bose condensed atoms confined in a harmonic trap with axial symmetry (\ref{extpot}).

Let us first focus our attention on the regime of low temperature.
Then, we solve numerically the BdG equations (\ref{B1:td})-(\ref{B4:td}) 
which are generally valid only in the limit $I \rightarrow 1$ as we have stated above (see details in Appendix).
Simulation parameters are summarized as : $\lambda= \sqrt{8}$ and $a/l_0=3.37\times 10^{-3}$ \cite{Jin1}.

\begin{figure} [ htb] 
  \centering {
  \includegraphics[scale=1]{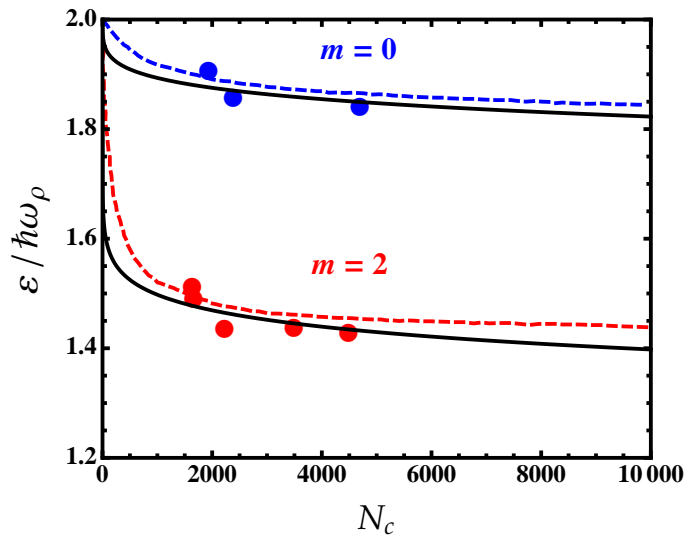}}
  \caption{ Frequencies of the  $m=0$ (top) and (bottom) $m=2$ modes as a function of the number of particles $N$.  
Solid lines are the predictions of TDHFBdG (\ref{B1:td})-(\ref{B4:td}). Dahsed lines are the predictions of the standard BdG \cite{Franc}.
Solid circles: experimental data of Ref.\cite{Jin1}.}
  \label{TES0}
\end{figure}

In figure \ref{TES0}, we compare our results for both $m=0$ and $m=2$ modes with the predictions of the standard BdG approximation (with $\tilde n=\tilde m=0$) \cite{Franc}
and the experimental data of \cite{Jin1}.
We see that small corrections to the standard BdG results \cite{Franc} brought our predictions in good agreement with the experimental results of \cite{Jin1}
revealing the important role played by the normal and anomalous correlations even at very low $T$.
The shift increases with increasing the interaction parameter $N_c a/l_0$ which is in fact natural since the anomlaous density itself grows with the interactions 
\cite{Boudj3, Boudj4,Griff}.
In the absence of interactions, the curves of the TDHFB approach and standard BdG predictions \cite{Franc} agree with each other 
and give $\varepsilon= 2 \hbar\omega_{\rho}$ for both modes.
 
\begin{figure}  [ htb] 
  \centering {
  \includegraphics[scale=1, angle=0]{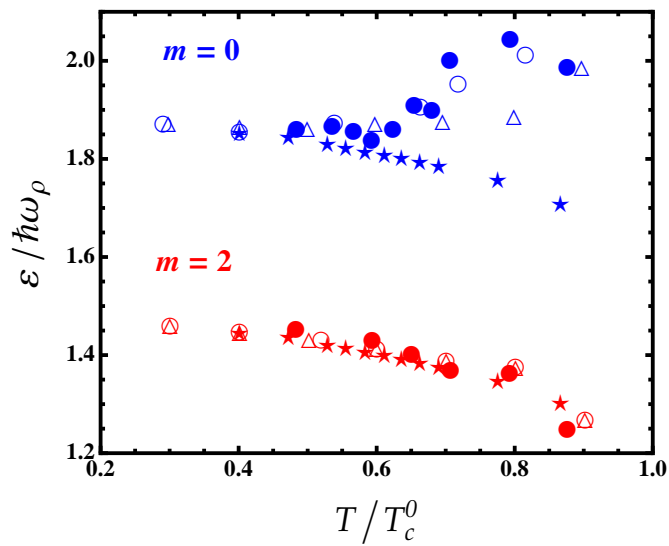}}
  \caption{ Frequencies of the  $m=0$ (top) and (bottom) $m=2$ modes as a function of reduced temperature $T/T_c^0$.  
Solid circles: experimental data of Ref.\cite{Jin1}. Open circles: TDHFB predictions. Stars: GHFB approximation \cite{Hut1}. Triangles: ZNG results \cite{Zar3}.}
  \label{TES}
\end{figure}

We turn now to study the collective modes of the whole system at any  nonzero temperature. 
To this end, we have to solve numerically the full TDHFBdG equations corresponding to (\ref{cond})-(\ref{Inv}) (see details in Appendix). 
As is depicted in figure \ref{TES},  for the $m=0$ mode, our solutions reproduce the JILA experiment \cite{Jin1} and other theoretical \cite{Hut1, Zar3} 
results quite well at temperatures up to $T \simeq 0.6 \,T_c^0$.
At $T > 0.6\,T_c^0$,  the TDHFBdG findings improve those obtained from the ZNG methods and 
successfully explain the anomalous behavior of the $m= 0$ mode.
This latter which produces in the excitation spectrum at higher temperatures ($T >0.6 T_c^0$),
is indeed originated from the inclusion of the dynamics of the thermal cloud with its normal and anomalous components.
In the same figure, on the bottom, we find good agreement with available experimental measurement and all existing theories for the $m=2$ mode
in the full temperature range.

\section{Conclusions and discussions} \label{concl}

In this paper we have studied the dynamics of quantum-degenerate Bose gas at finite temperature using the TDHFB theory.
The outcome of the numerical simulations of our equations shows a decrease of the condensed fraction and the critical temperature 
of an interacting Bose gas confined in an anisotropic harmonic trap with respect to the predictions of the HFB-Popov, QMC simulation and the noninteracting models.
This discrepancy can be attributed to the effect of anomalous fluctuations.

Most of our discussion has been devoted to the TDHFBdG equations.
Unlike other approaches, our equations allows us to study the collective modes of the condensate and the thermal cloud at any nonzero temperature. 
The solution of the TDHFBdG  equations in a spatially uniform gas provides a potentially useful expression for the excitation spectrum similar in the form to that of a binary condensate.
Higher order corrections to first and second sound velocities, thermal depletion and the superfluid fraction 
due to the finite temperature and the anomalous correlation are also precisely obtained.
It is worth noting that in the local density approximation \cite{Stamp} (LDA), the sound velocities, and therefore the excitation spectrum (\ref{dis}), 
can be considered to be defined at each point ${\bf r}$ in the BEC, where $c^c ({\bf r})=\sqrt{n_c ({\bf r}) g_{ca}/m}$ and $c^n({\bf r})=\sqrt{\tilde {n} ({\bf r}) g_{ac}/m}$. 
This approximation is valid as long as the Thomas-Fermi (TF) radius of the condensate in the  $k$-direction is much larger than the wavelength of the excitation 
\cite{Stamp, String1, Boudj11}. 
In the case of small noncondensed and anomalous fractions and for the parabolic density profile of the condensate in the TF regime,
one can expect that the excitation spectrum for the entire system in the LDA agrees 
well with the Bogoliubov spectrum measured  by Steinhauer et \textit {al}.  \cite{Stein} using the Bragg spectroscopy.

Additionally, we have extended our TDHFBdG equations to calculate the collective modes of a trapped atomic Bose gas in an anharmonic trap
and compared our result with the recent experimental measurement and previous theoretical predictions.
At temperature tending to zero, we found that our predictions correct the BdG results of Ref \cite{Franc} and well coincide 
with the experimental data of Ref.\cite{Jin1} for both modes $m=2$ and $m=0$. 
While all the theoretical treatments practically agree with each other in the spectrum of the non-interacting regime for both modes.

On the other hand, we have investigated the temperature dependence of the collective modes.
We showed that our findings excellently reproduce the experimental results for both the $m=2$ and $m=0$ modes. 
We have notably successfully explained the anomalous behavior of the $m= 0$ mode.
Such an anomalous behavior is in fact caused by the neglect of the dynamics of the normal and anomalous densities.


An interesting future work is to investigate theoretically the formation of the so-called thermal vortex (occurs during the condensation process) \cite {Ander}
employing our TDHFB model. 
Another topic of great relevance is the study of first and second sound in BEC with spin-orbit coupling at finite temperature.

\section*{Acknowledgements}
We are indebted to Werner Krauth  for providing us with the quantum Monte Carlo data.
AB would like to thank the LPTMS, Paris-sud for a visit, during which part of this work has been done.


\section*{Appendix: Numerical implementation} \label{A}

In this appendix we summarize how one would solve the TDHFB equations for an anisotropic trap.
Since the TDHFB equations are nonlinear coupled equations, they require a self-consistent and iterative approach to solve them. 
Finite-difference method has already been used to solve the TDHFB equations in order to analyze the vortex states at finite temperature \cite{Boudj6,Boudj7}.

Here, we use a basis-set method consisting of eigenstates of the trap potential \cite{Hut}.  
To find the full solution, one has to follow a series of actionable steps:
\begin{description}
  \item1. We solve Eqs. (\ref{cond})  and (\ref{anom}) or equivalently  (\ref{nonc})  for $\Phi (r)$ and $\tilde m (r)$ using the basis-set method.
  \item2. The condensed and the anomalous densities and the chemical potentials found
in step 1 are used to calculate both the spatially dependent effective interaction from (\ref{RCC}) and the excitation energies 
from equation (\ref{B1:td})-(\ref{B4:td}).
A noncondensed density $\tilde n$ is then updated from Eq.(\ref {heis}). 
New values for number of particles in and out of the condensate are derived from Eq.(\ref{Nprt}). 
  \item3. We solve the TDHFBdG equations in the basis of the eigenfunctions of the trap 
expanding the amplitudes $f_j^{c/n} (r)$ and $f_j^{c/n} (r)$ in the trap eigenfunctions \cite{Hut}.
In this basis we only need to calculate numerically the matrix elements of $g_{ca} n_c (r)$, $g \tilde n (r)$,  $g n_c (r)$, $g_{ac} \tilde n (r)$ 
in harmonic oscillator eigenstates.
  \item4. Iterate the above procedure until self-consistency is reached.

\end{description}

The algorithm can be checked by reproducing the ideal gas results where the problem reduces to 
the solution of the stationary Schr\"odinger equation for an anisotropic harmonic oscillator. 
In this case the anomalous density vanishes ($\tilde m (r)=0$)  \cite{Griff} and the condensate wavefunction reads
$\Phi ({\bf r})= \lambda^{1/4} \pi^{-3/4} \exp{[-(r^2+\lambda z^2)/2}]$.
For strongly interacting system i.e. $N \gg 1$, it is also possible to compare the numerical results with the analytic TF solutions. 

In our calculation, we found that a grid of $60 \times 60$ points covering the interval $0< r< 6$ (the same for $z$) is sufficient to describe the system
in the $(r,z)$ coordinates. Although the convergence of the ground state depends on the system parameters, this grid ensures convergence
with high accuracy.

\end{document}